# On The Organization Of Human T Cell Receptor Loci.

Amir A. Toor MD, [1] Abdullah A. Toor, [2] Masoud H. Manjili PhD, DVM. [3]

[1] Bone Marrow Transplant Program, Department of Internal Medicine, and [3] Department of Microbiology and Immunology, Virginia Commonwealth University, Richmond, VA; [2] Deep Run High School, Glen Allen, VA.

* Address correspondence to: Amir A. Toor, MD, Associate Professor of Medicine, Massey Cancer Center, Virginia Commonwealth University, E-mail: atoor@vcu.edu, Phone: 804-828-4360. 1300 E Marshall St., PO box 980157, Richmond, VA 23298

Word count: Abstract 327; Manuscript 6174;

Figures 6; Supplementary Tables 2; Supplementary Figures 2

**Abstract.**

The human T cell repertoire is complex and is generated by the rearrangement of variable (V), diversity (D) and joining (J) segments on the T cell receptor (TCR) loci. The T cell repertoire demonstrates self-similarity in terms clonal frequencies when defined by V, D and J gene segment usage; therefore to determine whether the structural ordering of these gene segments on the TCR loci contributes to the observed clonal frequencies, the TCR loci were examined for self-similarity and periodicity in terms of gene segment organization. Logarithmic transformation of numeric sequence order demonstrated that the V and J gene segments for both T cell receptor $\alpha$ (TRA) and $\beta$ (TRB) loci were arranged in a self-similar manner when the spacing between the adjacent segments was considered as a function of the size of the neighboring gene segment, with an average fractal dimension of ~1.5. The ratio of genomic distance between either the J (in TRA) or D (in TRB) segments and successive V segments on these loci declined logarithmically with a slope of similar magnitude. Accounting for the gene segments occurring on helical DNA molecules in a logarithmic distribution, sine and cosine functions of the log transformed angular coordinates of the start and stop nucleotides of successive TCR gene segments showed an ordered progression from the 5' to the 3' end of the locus, supporting a log-periodic organization. T cell clonal frequencies, based on V and J segment usage, from three normal stem cell donors were plotted against the V and J segment on TRB locus and demonstrated a periodic distribution. We hypothesize that this quasi-periodic variation in gene-segment representation in the T cell clonal repertoire may be influenced by the location of the gene segments on the periodic-logarithmically scaled TCR loci. Interactions between the two strands of DNA in the double helix may influence the probability of gene segment usage by means of either constructive or destructive interference resulting from the superposition of the two helices.

**Introduction**

T cells are central to the normal execution of adaptive immunity, allowing identification of a multitude of pathogens and transformed cells encountered in an organism's lifetime. T cells accomplish this task by recognizing peptide-MHC complexes by means of hetero-dimeric T cell receptors (TCRs) expressed on their surface. The TCR serve the primary antigen recognition function in adaptive immune responses. TCRs are comprised of either, an alpha and a beta chain (TCR $\alpha\beta$) in the majority of T cells, or less frequently, gamma and delta chains (TCR $\gamma\delta$). [1] The ability of the human T cells to recognize a vast array of pathogens and initiate specific adaptive immune responses depends on the versatility of the TCR, which is generated by recombination of specific variable (V), diversity (D) and joining (J) segments in the case of TCR $\beta$, and unique V and J segments for TCR $\alpha$, $\delta$ and $\gamma$. Complementarity determining regions (CDR) are the most variable part of the TCR and complement an antigen-major histocompatibility complex's shape. The CDR is divided into 3 regions termed CDR1-3, and of these CDR1 and CDR2 are coded for entirely by the V segment, whereas CDR3 incorporates a part of the V segment and the entire D and J segments for TCR $\beta$ and parts of the V and J segments for TCR $\alpha$. CDR3 is the most variable region and interacts with the target oligo-peptide lodged in the antigen binding groove of the HLA molecule of an antigen presenting cell [2]. The germ line TCR $\beta$ locus on chromosome 7q34 has two constant, two D, fourteen J and sixty-four V gene segments, which are recombined during T cell development to yield numerous VDJ recombined T cell clones; likewise, TCR $\alpha$ locus on chromosome 14q11 has one constant, sixty-one J and fifty-four V segments [3]. Further variability and antigen recognition capacity is introduced by nucleotide insertion (NI) in the recombined TCR $\alpha$ and $\beta$ VDJ sequences. This generates a vast T cell repertoire, yielding in excess of a trillion potential TCR$\alpha\beta$ combinations capable of reacting to non-self (and self) peptides. Since the advent of next generation sequencing (NGS) techniques, the TCR repertoire, as expressed by TCR $\beta$ clonal frequency has revealed that the T cell repertoire in healthy individuals is complex with thousands of clones in each individual spanning a spectrum of high and low frequencies [4, 5].

T cells have a fundamental role in clinical medicine, especially in cancer therapeutics. As an example, clinical outcomes following stem cell transplantation (SCT) are closely associated with T cell reconstitution, both from the standpoint of infection control and control of malignancy [6,7]. A detailed examination of transplant outcomes in the context of immune reconstitution suggests that the T cell recovery is a dynamic and ordered progression of events, which follows mathematically quantifiable rules. In other words, T cell reconstitution over time following SCT may be considered as a dynamical system, where each of the successive states of the system (T cell repertoire complexity) when plotted as a function of time can be described mathematically as a non-random process [8, 9]. Consistent with this hypothesis, T cell repertoire has been shown to posses a fractal self-similar organization with respect to TCR gene segment usage [10].

Clinical outcomes following SCT have been studied using probability theory, assuming that the outcomes are inherently random within specific constraints, such as, HLA-matching and intensity of immunosuppression. However, the dynamical system model of immune reconstitution implies that T cell repertoire generation, and thus clinical outcomes tied to it, is not a stochastic process but a mathematically definable one. Since the T cell repertoire is dependent on the clones generated by TCR gene rearrangement, it may be postulated that the TCR locus may exhibit organizational features reflected in the resulting T cell repertoire. It has been observed that the T cell clonal repertoire follows power laws and demonstrates self-similarity across scales of measurement, which are features of fractals. Fractal geometry represents iterating functions, which result in structures demonstrating organizational self-similarity across scales of magnitude. Mathematically, self-similarity is characterized by the logarithm of magnitude of a parameter ($y$) maintaining a relatively constant ratio to the logarithm of a scaling factor value ($x$), this ratio is termed fractal-dimension (FD) [11]. FD takes on non-integer values between the classical Euclidean integer dimensional values. Self-similarity is a structural motif widely observed in nature, albeit on a limited scale, such as in the branching patterns of trees, or of vascular and neuronal networks in animals [12, 13, 14, 15]. Fractal organization has also been described in terms of molecular folding of DNA, and in nucleotide distribution in the genome [16, 17, 18, 19]. In such instances fractal dimension explains the complex structural organization of natural objects. A germ-line linear strand of DNA such as the TCR locus, rearranged to yield new complex patterns, may be similarly considered. In this case the recombination of TCR V, D and J segments, and NI lends complexity to the rearranged locus compared to its native state. The resulting T cell repertoire with its many unique T cell clones may then be considered to have a fractal organization, if the frequency of T cell clones with unique D, DJ, VDJ and VDJ(+NI) segments are considered sequentially at an increasing level of clonal definition. Using such analyses the T cell repertoire appears highly ordered with respect to TCR β VDJ gene segment usage demonstrating self-similarity (in terms of clonal frequency) at multiple levels of clonal definition. When the number of TCR segments used to define a clone is used as the scaling factor, a consistent fractal dimension value of 1.6, 1.5 and 1.4 is observed across normal stem cell donors when TCR clone families bearing unique J, VJ and VJ+NI are evaluated [10]. It stands to reason that the TCR loci which generate this T cell clonal diversity should be organized in a similar manner to result in a repertoire so ordered.

The universal constants π and *e* have a role in defining the fractal dimension of the T cell repertoire, which would then imply that the TCR locus may be arranged in conformity with these constants, and posses self-similarity as well as periodic characteristics. An inspection of the known sequence of the TCR loci with attention to V and J gene segment size and spacing supports this notion. In this paper, an examination of the TCR α and β loci, in light of *e* and π is presented supporting the hypothesis that TCR locus organization may influence T cell repertoire. The findings reported here support the hypothesis that the T cell repertoire is not a randomly generated set of T cell clones, rather it is the result of a deterministic process dependent on TCR gene segment distribution on the individual loci.

**Methods.**

*Human T cell receptor loci*

Data on the TCR α/δ (TRA) and β (TRB) loci were obtained from the NCBI, using the public database, (http://www.ncbi.nlm.nih.gov/nuccore/114841177?report=graph; http://www.ncbi.nlm.nih.gov/nuccore/99345462?report=graph) (Supplementary Tables 1 and 2). The graphic format, with locus identification was utilized. Data examined included the position of the initial (start) and final (stop) nucleotides of all the TCR gene segments, beginning from the 5' end of the locus and going to the 3' end. Segment length was taken from the NCBI data-base, and spacing between consecutive segments was calculated by taking the difference between the numeric value of the final nucleotide or base pair  ($x_f$) position of a gene segment and the initial nucleotide position ($x_i$) of the following segment (Supplementary table 1 and 2). In the ensuing calculations, numeric data were either considered without any modification or transformed to natural logarithms to eliminate the effect of relative magnitude between the variables being examined, and also to allow comparisons across different scales, e.g. segment size in hundreds or tens of nucleotides versus inter-segment spacing in the thousands or hundreds of nucleotides respectively. Microsoft Excel (version 14.2.5) software was used to perform various calculations presented in this paper.

*Gene Segment Distribution on the TCR Locus*

To investigate the possible fractal nature of the TCR locus, analysis of gene segment distribution was performed utilizing the relationship between magnitude and scaling factor across varying levels of magnification. The resulting fractal dimension (FD) is described as follows

$$FD = Log\ (magnitude)\ /\ Log\ (scaling\ factor)$$

To determine the fractal dimension of the TCR loci, the FD was calculated analogous to the calculation of the fractal dimension of a Von Koch curve; the length of the gene segment in number of nucleotides or base pairs was considered as a scaling factor and the length of the inter-segment space, also in nucleotide number, as the magnitude of the line segment. This calculation was performed individually for each segment to fully account for variability observed.

*TRB locus Periodicity.*

The TCR gene segments occur periodically from the 5' to the 3' end of the loci, with V, D in TRB, J and C segments in that order. For the calculations regarding the gene segment periodicity and its influence on gene usage frequency, the helical DNA molecules were considered as a propagating spiral (or a wave). In this model each base-pair on a strand of DNA may be considered as a point $x$, with subsequent base pairs, $x+1..., x+n$ being successive points on the spiral, as opposed to points on a straight number line. The spiral or helical DNA molecule, as it executes one turn goes through ~$2\pi$ radians, in terms of angular distance spanned. One turn of the helix contains 10.4 nucleotides [20], so the space between successive nucleotides may be considered as angular distance, in radians between them (assuming a uniform unit radius of the DNA molecule). This inter-nucleotide 'distance' will be $2\pi/10.4$ (Supplementary Figure 1). The spatial position of any nucleotide $x$, relative to the locus origin may be then be described as the angular distance in radians (($2\pi x/10.4$) radians) and its coordinates on the DNA molecule determined.

*T cell clonal frequency*

Stem cell donor samples for determining T cell clonal frequency were obtained as part of a clinical trial approved by the institutional review board at Virginia Commonwealth University (ClinicalTrials.gov Identifier: NCT00709592). As previously described, CD3+ cells were isolated from stem cell transplant donor samples and cDNA synthesized from these cells [10]. The cDNA was then sent to Adaptive Biotechnologies (Seattle, WA) for high-throughput sequencing of the TCR β CDR3 region using the ImmunoSEQ assay. This approach is comprised of a multiplex PCR and sequencing assay in combination with algorithmic methods to produce approximately 1,000,000 TCR β CDR3 sequences per sample [21]. The assay utilized 52 forward primers for the Vβ gene segment and 13 reverse primers for the Jβ segment, to generate a 60 base pair fragment capable of identifying the entire unique VDJ combination [22]. Amplicons were then sequenced using the Illumina HiSeq platform, and data was analyzed using the ImmunoSEQ analyzer set of tools. This approach enables direct sequencing of a significant fraction of the TCR repertoire as well as permitting estimation of the relative frequency of each CDR3 region in the population.

**Results.**

*Self-similarity of the TCR loci*

Sequences of the TCR α/δ (TRA) and β (TRB) loci were examined and the position of the initial (start) and final (stop) nucleotides of the TCR gene segments, beginning from the 5' end of the locus and going to the 3' end were recorded. Segment length was taken from the NCBI data-base, and spacing between consecutive segments was calculated (Supplementary Tables 1 and 2). Self-similarity across the T cell receptor α and β loci was first examined by deriving the fractal dimension of these loci. To accomplish this, the spacing of all the V and J segments across the α/δ and β loci was determined, in numbers of nucleotides along the DNA molecule, and the natural logarithms of these values calculated. Given the relative uniformity of sizes (in nucleotides) for the V and J gene segments, these were considered as a scaling parameter for determining the fractal dimension of the TCR loci (FD-TCR). The sizes of neighboring segments and intersegment spaces were used in the FD-TCR calculations to avoid bias. The following formula was used

$$FD\text{-}TCR = Log \text{ (Intron length following } n^{th} \text{ TCR seg.) } / Log \text{ (Length } n^{th} \text{ TCR seg.)} \quad \dots \text{ [1]}$$

Relatively consistent values of FD-TCR were observed across the distribution of the V and J segments for both the TCR α and β loci (Figure 1), when the calculated fractal dimensions were plotted across the loci. This indicates that when viewed on a logarithmic scale there is uniformity in the distribution of gene segments both within and between the different TCR loci, a hallmark of self-similar systems. The average FD-TCR of the TRB V and J segments were 1.4±0.1 and 1.3±0.2 respectively. Corresponding values for TRA locus were 1.5±0.1 and 1.7±0.1, for the V and J segments. The self-similarity across the TRA locus may also be seen when the spacing between successive gene segments is plotted from the 5' to 3' end in a circular area graph (Figure 2). The two halves of the figure are similar in appearance and are symmetric, except for the spacing between TRA-J segments being an order of magnitude lower when compared with TRA-V segments. Notably, oscillation around a central value is observed in these graphics, suggesting that periodic variation is present in the measured parameters. Despite this variability, the narrow distribution of the FD-TCR values was consistent with self-similarity in the organization across the V and J segments of the two TCR loci examined. This implies that the size of the interval between successive T cell gene segments is proportional to the size of the gene segment. It may be postulated that this phenomenon exerts an influence on the order of TCR gene rearrangement, specifically the Dβ to Jβ and DJβ or Jα to Vβ or Vα recombination.

In calculating FD-TCR, the D and C segments were not considered because of their infrequent and non-periodic occurrence, as well as their being, interspersed between other loci. However, notably their size followed fixed proportion to the sizes of J and V segments respectively, such that D segments were ~1/3 to 1/4 the size of J, and the C segments were approximately 3 times the length of the V segments. Similarly the TCR-γ locus was also not evaluated because of the small number of gene segments, however, it is to be noted that the gene segment length was similar to the TRA and TRB loci.

*Logarithmic Distribution of V segments with respect to D and J segments*

During T cell receptor recombination, the J and the D segments in TRA and TRB loci respectively, are recombined with the V segments. The distribution of V segments in the TRA and TRB loci was therefore examined relative to the position of the J and the D loci respectively, to discern if any organizational order was evident. The relative position of V loci was calculated with respect to the two D loci for TRB, and selected J loci for TRA, by the using a formula considering the 5'-initial nucleotide positions ($x_i$) of the D or J gene segments, and the final, 3' nucleotide position of the V segments ($x_f$) from the origin of the locus

*Relative recombination distance $V_n = x_i$-D <u>or</u> J segment / $x_f$ $n^{th}$ V segment*        ……                [2]

Relative recombination distance (RRD) demonstrated a consistent proportionality in the distribution of V segments across both TRA and TRB loci. When RRD for successive TRB V loci was plotted in order of occurrence on the locus, the value declined as a logarithmic function of V segment rank, with a slope of 1.6 for the TRB locus (Figure 3A) and ~1.3 for the TRA locus (Figure 3B & 3C), values very close to the FD-TCR. These calculations demonstrate that the V segments are spaced from their respective D and J segments in a logarithmically proportional manner. This is consistent with the self-similar nature of the TCR loci as seen in the preceding calculations and also validates the notion that the distribution of various gene segments on the TRA and TRB loci is not random, but rather follows a mathematically determined order, apparent when logarithmic transformation is used.

*Logarithmic scaling of the TCR gene segment periodicity*

Repetitive occurrence of gene segments on the TCR loci, as well as the fluctuation of FD-TCR about the average value (Figure 1) suggested that they conform to a periodic distribution analogous to the periodic behavior exhibited by phenomenon such as wave motion, or in this case helix/spiral progression. To examine the periodicity of the relative positions of gene segments on the TCR loci they were considered as successive nucleotide sequences on the DNA helix and the angular distance between segments determined by using the relationship *$2\pi x/10.4$*, where *x* is the initial or final nucleotide position in a gene segment with respect to the TCR locus origin (Supplementary Figure 1). The calculated angular distance between the gene segments were further analyzed by determining the distance between V and D segment in TRB.

These values were then used to determine the coordinates of the gene segments on the DNA helix, using the trigonometric parameters, sine and cosine for the initial nucleotides ($x_i$) relative to the locus origin. When this was done for the angular distance, *$2\pi x/10.4$*, and the resulting sine and cosine values plotted against the angular distance (*$f(x) = Sin(2\pi x/10.4)$* or *$Cos((2\pi x/10.4)$*) from locus origin no clear pattern was observed, with the sine and cosine values for each of the positions distributed randomly along the DNA strand (Figure 4A). Given the previously observed logarithmic ordering of the TCR loci, the trigonometric functions of the natural logarithm of 5'-initial nucleotides angular distance from the locus origin (*$Sin(Log(2\pi x/10.4)$* and *$Cos(Log(2\pi x/10.4))$*), were then plotted against the angular distance. This demonstrated that the TCR gene segments were positioned on the locus in a sequential aligned manner (Figure 4B) for both the TRA and TRB loci when the log-periodic functions of the nucleotide positions are considered. Further, when the TRA and TRB sine and cosine plots were overlaid, these plots were superimposable, supporting the earlier observation regarding the self-similar nature of these loci (Figure 4C). These calculated trigonometric functions of TCR gene segments reveal that these loci have a log-periodic nature.

*T cell clonal frequency and its correlation with TRB gene segments*

The periodic nature of the TCR gene segments on the locus from an evolutionary standpoint may be the result of TCR locus generation being an iterative process, which would be a logical conclusion given its role in adaptive immunity of an organism. To determine whether the periodicity of the TCR loci was reflected in the T cell clonal frequency, high throughput TRB sequencing data from 3 normal stem cell transplant donors was used. Clonal frequency for unique TRB V and TRB J segments was plotted against the angular distance (in radians) between the 5' initial-nucleotide ($2\pi x_i / 10.4$) of the TRB-D1 (& TRB-D2) and the 3' final-nucleotide ($2\pi x_f / 10.4$) of the V segments or the 5' start nucleotide of the J segments. The T cell clonal frequency varied as a quasi-periodic function of the angular distance between both TRB-D1 and D2 and the successive TRB-V segments, oscillating between high and low clonal frequency values (Figure 5A). For the TRB-V clonal families, the peaks of clonal frequency occurred approximately every 50,000-60,000 radians going in the 5' direction from the D loci. Further, the two J segment bearing areas of the TRB locus were approximately 5000 radians apart (in the 3' direction) and demonstrated oscillating clonal frequency (Figure 5B). This finding was consistent between unrelated stem cell donors, demonstrating very high expression levels for some loci, intermediate for others and low or no expression in others (Supplementary Figure 2). This may be interpreted as TCR gene V segment functional recombination *probability amplitudes* oscillating between 0 (no recombination) and 1 (very frequent recombination resulting in high clonal frequency) across the locus.

**Discussion.**

The germ-line genomic T cell receptor (and immunoglobulin) loci have the unique characteristic in that these can undergo DNA double strand break and recombination resulting in VDJ rearrangement which results in the generation of numerous unique T cell clones with the ability to identify the wide array of antigens. In recent years new insights have been gained into the mechanism of this very ordered sequence of DNA rearrangement. This process results in a very large repertoire of immune effectors, which is employed therapeutically in modalities such as SCT. Clinical outcomes following transplantation in particular, and immune therapies in general have been modeled as stochastic processes. However, recently a dynamical system model for immunological responses has been proposed. This model postulates that immune responses are not stochastic, but are predictable and accurately quantifiable as they evolve over time. In support of this, T cell repertoire has been recognized as possessing characteristics of self-similar systems, and following power law distributions. If the T cell repertoire is non-random and describable in mathematical terms, it is logical that T cell gene rearrangement contributes to the emergence of this order. This would imply that the TCR locus organization lends itself to a mathematically quantifiable gene rearrangement process. In this manuscript, quantitative uniformities are explored across the TCR loci, and their potential role in shaping the T cell repertoire investigated.

T cell receptor rearrangement is a strictly ordered process, with the sequence of recombinatorial events governed by a numeric, so-called '12/23 rule'. Considering the example of the T cell receptor β, the V, D and J segments are flanked by conserved sequences called recombination signal sequences (RSS), comprised of a heptameric and a nonameric sequence, which are interposed by either a 12 (RSS-12) or a 23 (RSS-23) base pair sequence. VDJ rearrangement is brought about by recombinase-activating gene-1 (RAG-1) and RAG-2 protein complexes, which always bring together segments with a RSS 12 with a segment flanked by a RSS-23, not otherwise. Considering the TRB; Dβ segments are flanked by the RSS sequences on both sides (5', RSS-12 and 3', RSS-23), Jβ segments on the 5' end (RSS-12) and Vβ segments on the 3' end (RSS-23) [23]. In the series of events set off during T cell development, initially a Dβ segment rearranges with one of the Jβ segments, then, the combined DJβ joins with a specific Vβ segment to result in VDJβ rearrangement, yielding a unique T cell clone. Similar considerations hold for TCRα, where Jα recombines with Vα, with the additional possibility of locus editing such that alternative 5' Vα may be rearranged to an alternative 3' Jα segment at a later time [24]. As elegant as this system is, it does not completely explain either, how the order of recombination is determined or what determines the variability in the use of various TCR gene segments encountered in the T cell repertoire; for instance, why does the Dβ RSS-23 preferentially rearrange with Jβ RSS-12, and not the Dβ RSS-12 with the Vβ RSS-23, and, why are clones with TRB V5-1 encountered much more frequently than TRB V5-4 in the normal T cell repertoire? Further, in the TRD (TCR-δ) locus rearrangement under normal circumstances there is strict regulation such that TRA and TRD segments do not recombine with each other, despite the TRD locus being nested within the TRA locus. However, when a TRA-Jα segment is ectopically introduced in the TRD locus at the TRD-Dδ segment position, rearrangement of this ectopic TRA-J with TRD-V segments occurs, demonstrating that locus position, rather than the actual sequence may have a deterministic role in the order of recombination [25]. This raises the question of whether there are further structural motifs on the DNA molecule, which determine the order of recombination.  It is this question, that the log-periodic nature of the TCR locus elucidated in our analysis may help answer.

In nature there is a tendency for organizational patterns to be repeated over different scales of measurement and for such patterns to be observed across different systems. Fractal organization in the VDJ segment usage in the T cell repertoire of normal individuals has been observed with the diversity, joining and variable gene segment usage defining a virtual 'structure' that results from recombination of the T cell β receptor locus [10, 26, 27]. With this background the proportions between the V and J segment size and inter-segment intron lengths were examined relative to each other and were found to be similar. It is likely that the proportional distribution of V and J segment size and spacing between individual segments (fractal organization) in this instance serves to order the ensuing rearrangement process. In essence, the V segments are similar in their organization to the J segment, and from an evolutionary standpoint the V segment sequence represents a *logarithmically magnified* version of the J segment sequence. This may explain why in the order of gene segment rearrangement, Dβ to Jβ and DJβ or Jα to Vβ/α segments, RAG complexes are always directed from the shorter, closely spaced J segments to the longer, more dispersed V segments, such that the reverse does not transpire in the course of normal recombination. Further, the logarithmic scaling implies that the distribution of these size-ordered segments is always similar in their respective sections of the TCR locus, which ensures that RAG complexes do not have to 'scan' an entire sequence of nucleotides to randomly encounter a coding segment, but can potentially align with relevant segments, skipping over given lengths of intronic material. This would then provide an additional mechanism to complement the 12/23 rule, and to ensure fidelity of recombination. Presumably the epigenetic mechanisms such as RAG2 interacting with methylated of histone3-K4, further facilitates the navigation along the TCR loci to make the process a model of efficiency. Conformational changes in the locus, which bring V segments in apposition to J segments, may also be dictated by the logarithmic relationship. This hypothesis, if true, suggests that the origin of the fractal organization of the T cell repertoire is within the arrangement of the TCR loci resulting in an ordered recombination process. The log-periodic nature of other fractal phenomenon encountered in nature supports this postulate [28, 29].

High-throughput sequencing of TRB has demonstrated a differential representation of the different gene segments in the T cell clonal repertoire, indicating that some sequences are used at a higher frequency than others [3, 4, 10]. This has been observed for TCRγ as well as TCRβ and has been seen for both J and V segments [30]. This recombination bias affects both in-frame and out-of-frame recombined sequences, suggesting that it is not a consequence of thymic selection, nor HLA restriction, rather is a result of recombinatorial usage bias, or ranking of various segments. Figure 4, demonstrates this phenomenon, and it is also reflected in the power law distribution of the final T cell clonal distribution observed. The relationship between TCR locus organization and segment selection in this rearrangement process and its impact on the T cell repertoire generation has been a focus of intensive study in the recent years. Recently a biophysical model describing yeast chromosome conformation has been applied to the murine TCR β–D and -J segment and the derived model based on 'genomic distance' between these segments has partially recapitulated the observed bias in J segment usage [31]. This strengthens the notion that chromatin conformation has a formative role in the T cell repertoire generation. Regardless of the mechanism of recombination it has become obvious that the T cell repertoire that emerges has a 'biased' VDJ segment usage, with certain segments being used more frequently than others. This suggests that these segments may be more efficiently rearranged resulting in their over representation in the repertoire and vice versa. Could this in part, be a function of the intrinsic position of the segment on the DNA strand versus being entirely determined by extrinsic factors, such as epigenetic influence and segment localization on the chromosome?

Given the emergence of the constant π in the equations describing the fractal nature of the T cell repertoire in normal stem cell donors and the periodic nature of TCR gene segments on the TCR locus, their relative positions were examined using trigonometric functions to account for the helical nature of DNA. Similarity was observed in the relative location of the V, D and J segments across the TRA and TRB loci when they were examined using logarithmic scaling, with increasingly complex waveforms observed as higher order harmonics were evaluated (data not shown). There are several important implications of this observation. First, analogous to the phenomenon of superposition (constructive or destructive interference) observed in the mechanical and electromagnetic waves, one may consider that relative position of a particular segment, reflected by the sine and cosine functions and angular distance from the rearranging segment (Dβ or Jα), may influence its usage in repertoire generation resulting in the periodic distribution of the V and J segment usage in T cell clones when the locus is interrogated from the 5` to 3` end. Essentially, this means that using analytic techniques such as Fourier's series, *probability amplitudes* may be determined for the various gene segments on the TCR loci based on their positions. It may be very likely that the recombination is most frequent for gene segments that occur at a certain 'harmonic' frequency. As an example in the data presented the TRB-V segment clonal frequency appears to *oscillate* with a *wavelength* of ~50-60,000 radians from the TRB-D segment. It may be speculated that the gene segment distribution periods represent optimal energy distribution for recombination to occur on the long helical DNA molecule, analogous to the *interference* phenomenon encountered in wave mechanics. This is plausible because the DNA double helices may represent two superposed waves, and the gene segment location may lend itself to either constructive or destructive interference, impacting the interaction with RAG genes and recombination potential. This would in turn determine the probability amplitude of that gene segment being represented in the final T cell clonal repertoire (Figure 6).

In this hypothetical paper we demonstrate that the TCR loci have an iterative, logarithmically scaled periodic nature when the gene segment distribution is considered. This means that with minimal genomic material an organism may efficiently generate a vast receptor array, protecting the organism from populations of antigenically diverse pathogens it is likely to encounter in its lifetime. Secondly, nature has a tendency to be conservative in terms of using the same basic processes with minor variation in different settings to generate variability observed. The data presented here make it possible to consider variations in DNA expression as a partial function of the wave-mechanical properties of the DNA double helix. Further it may be postulated that if self-similarity is true for the distribution of gene segments within a locus than it is likely to be true for the genome as a whole, and the distribution of exons and introns in the genome may be similarly considered. The expression of these genes may then be determined by the sum of the harmonics of the relevant genes and be predictable using Fourier analysis. This work demonstrates that the T cell receptor loci, and by extension the T cell receptor repertoire is not an accident of nature, a result of random mutations, rather it is a mathematically ordered iterative process. In essence evolution of the immune system conferring a survival advantage to organisms in higher order phyla. These findings strengthen the argument that immune responses, such as following SCT represents an example of an ordered dynamical systems and not a stochastic process.

Acknowledgments: The authors gratefully acknowledge Ms. Kassi Avent, and Ms. Jennifer Berrie for technical help in performing the high throughput TRB DNA sequencing. The TRB sequencing was performed by Adaptive biotechnologies, Seattle, WA. Mr. Abdullah Toor collected the data and did most of the calculations reported in the paper. Dr. Amir Toor developed the idea and wrote the paper, as well as performing some of the calculations. Dr. Masoud Manjili supervised the TRB sequencing and critically reviewed the manuscript.

# Figures:

**Figure 1.** TCR α (TRA) and β (TRB) fractal dimension calculated for each variable and joining segment [Formula 1]. Values at each point are plotted along the nucleotide positions on the locus. TCR δ region on the TRA locus is excluded, as are the D and C segments for the TRB locus.

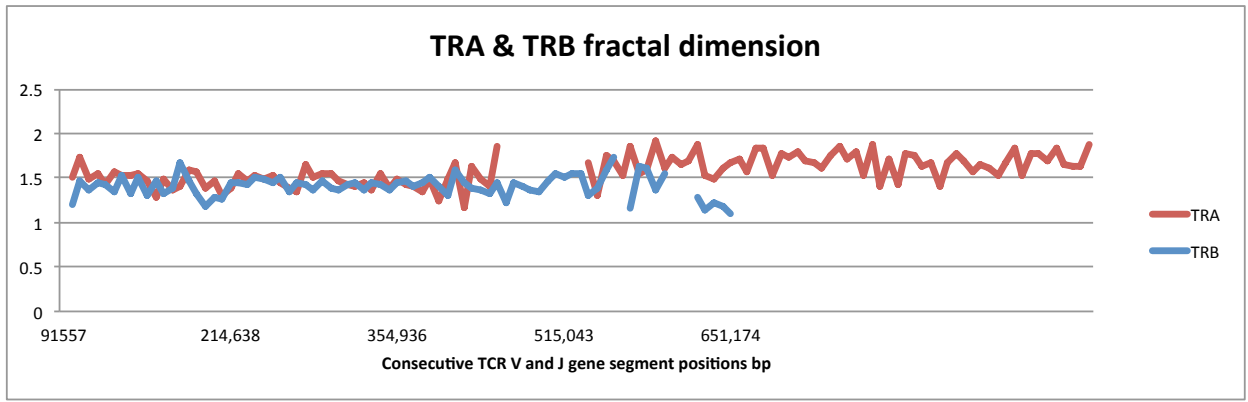

**Figure 2.** Relative position of the first nucleotide of each TRA gene segment from the 3' end (blue area) of the locus plotted against spacing (red area) following that TCR gene segments (Y-axis, Log₁₀-scale). This demonstrates self-similarity in the gene segment size and spacing distribution across the V (right) and J (left) loci, with the two halves of the figure demonstrating symmetry. Log scale used.

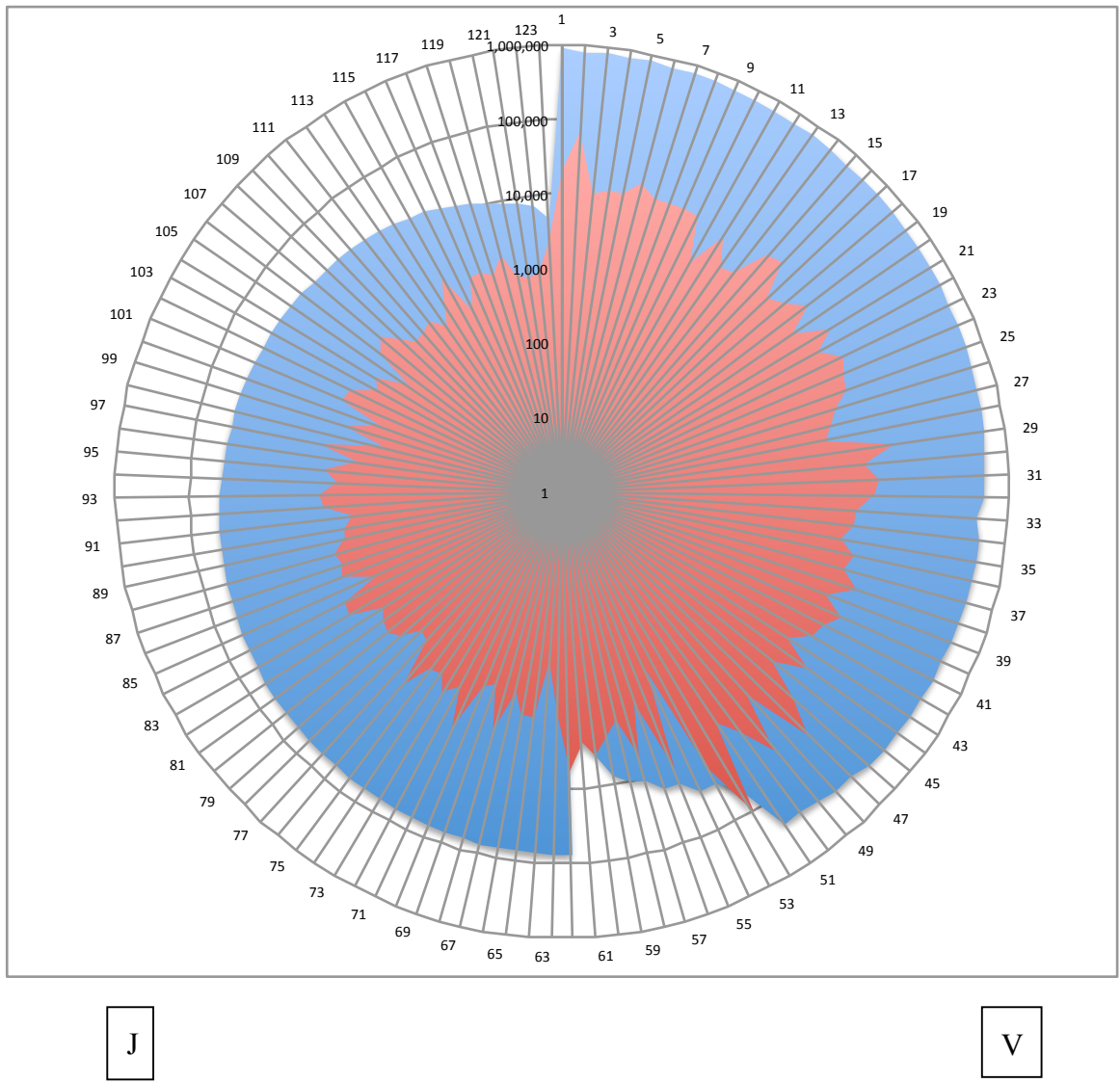

**Figure 3.** Relative recombination spacing [formula 2] for successive TRB-V segments with TRB-D1(A). Spacing for each V segment plotted in succession starting from V1 to V29. V30 excluded because of its location 3` to the D segments.  (B & C) RRS for each of the 40 TRA-V segments with the first (B) and last (C) J segment.

A

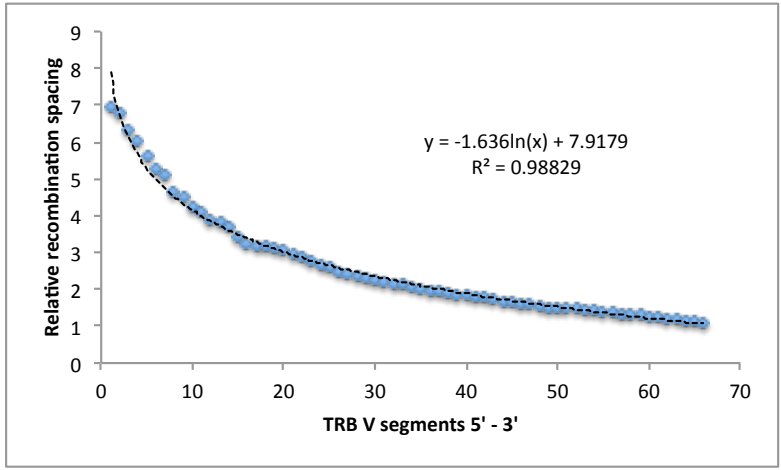

B

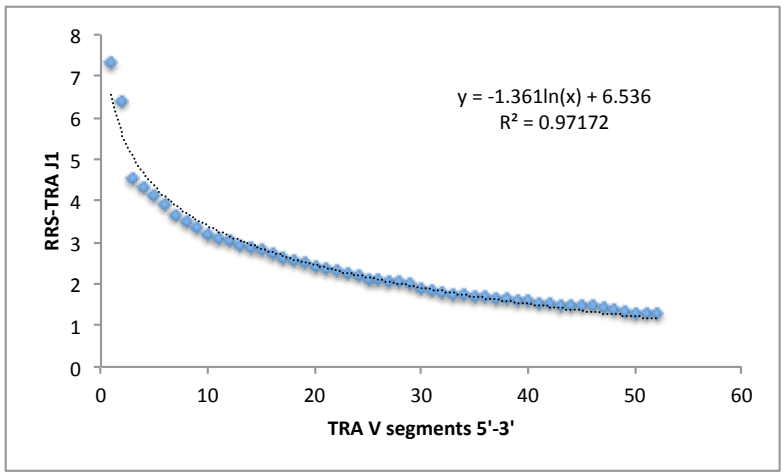

C

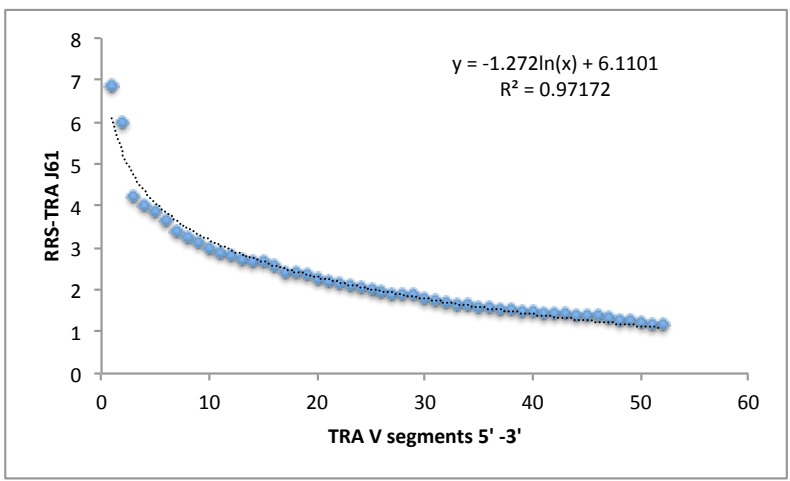

Figure 4. Logarithmic ordering of periodic TRB gene segments. (A) Sine (blue diamonds) and cosine functions (red) of TRB gene segment 5' initial nucleotide's angular distance from locus origin (5' end) plotted across the TCR loci.  (B) Sine (orange diamonds for TRB; blue for TRA) and cosine functions (blue for TRB; red for TRA) of the natural logarithm of TRB gene segment 5' first nucleotide and TRA 3' last nucleotide angular distance from locus origin (5' end) plotted across the TCR loci. (C) TRA and TRB sine and cosine coordinates superimposed, demonstrating self-similarity across the two loci.

A

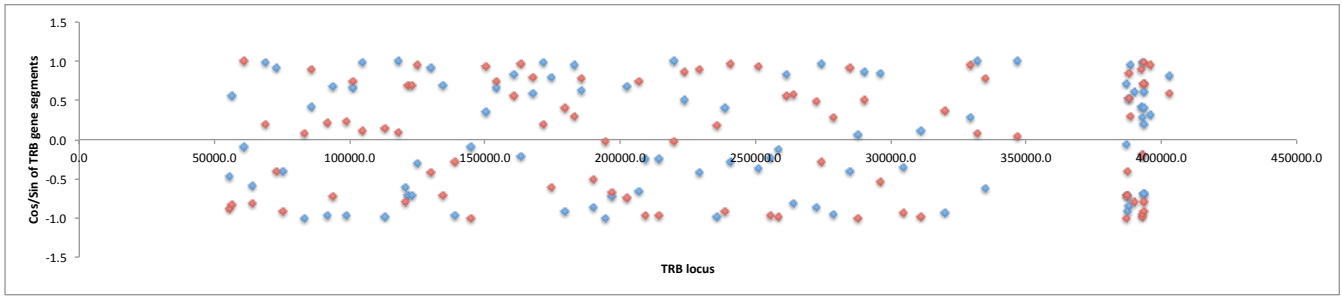

B

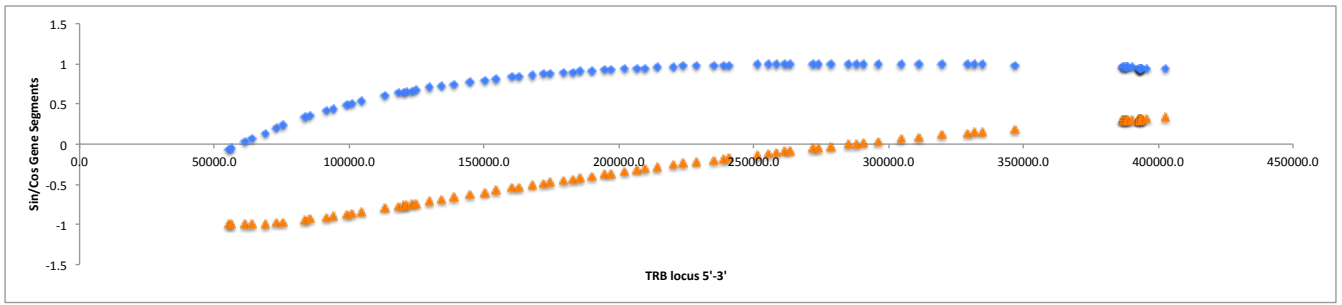

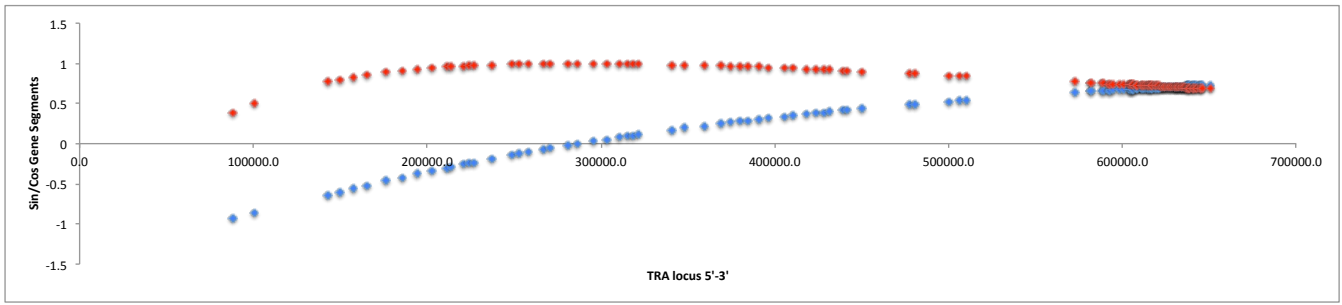

C.

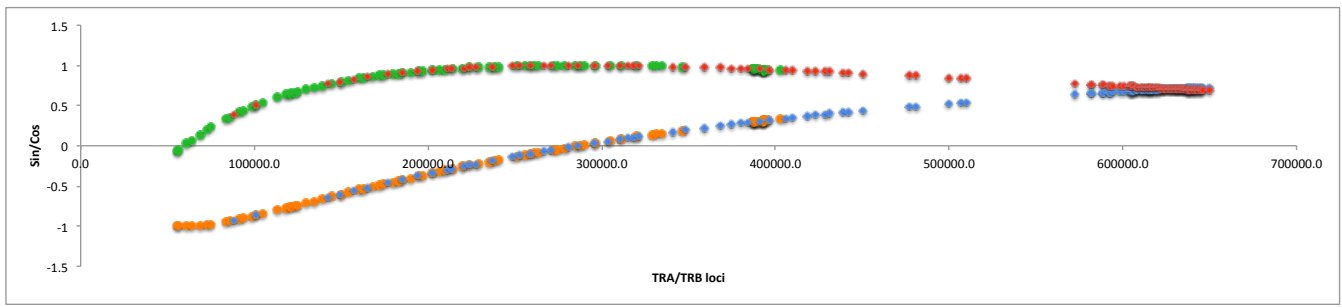

Figure 5. T cell clonal frequency as a periodic function of the V and J gene segment positions on the TRB locus. (A) TRB-V clonal frequency demonstrates quasi-periodicity (irregular) as a function of angular distance between TRB-D1 and successive TRB-V segments from V29 to V1. This represents periodic fluctuation in functional recombination probability amplitude across the locus. (B) TRB-J clonal frequency as a function of the angular distance between TRB-D1 and successive TRB-J segments. Both X and Y-axis magnitude different by approximately an order of magnitude (one-log) between the two graphs.

A

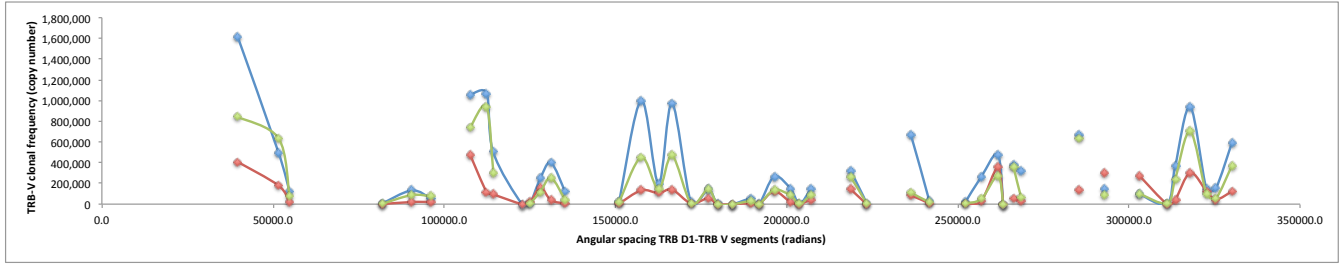

B

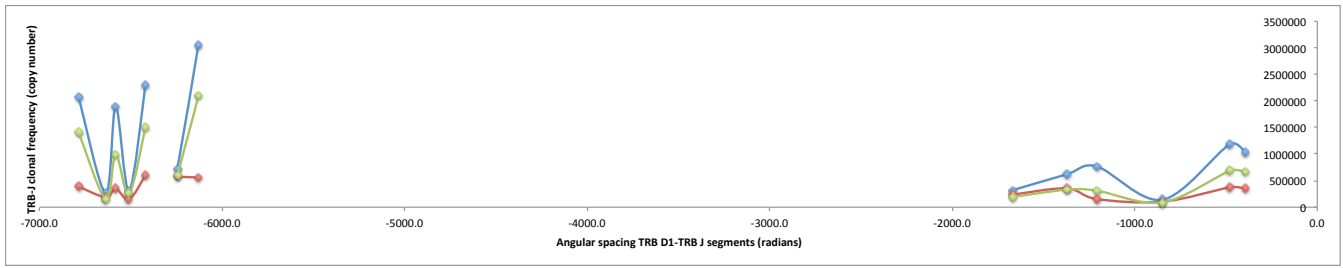

**Figure 6**. A model depicting T cell receptor organization and its influence on gene segment recombination probability. TCR V segments are separated by long intervals, J segments by shorter intervals (dashed lines); the ratio of *log* segment length to *log* spacing is approximately ~1.4 for V segments and ~1.3 for J segments. Relative interval between successive V segments and the J segments in the TRA locus (top blue curve), declines *logarithmically* with a slope of ~1.3. Sine and cosine function value of the start nucleotides of each V segment extrapolated to the sense (green) and antisense (blue) DNA strands, demonstrate that the gene segments are accurately aligned once the logarithmic organization is accounted for. Hypothetically the segment location on the two DNA helices being in-phase or out-of-phase may impact the energetics of DNA-RAG enzyme interaction and thus the *probability amplitude* (orange line, going from 0 to 1) for gene segment recombination analogous to wave *interference* phenomenon. In the model depicted, V1 location on the two helices is out of phase, V2 is partially in phase and V3 is completely in phase. Closely clustered together J segments are more likely to be in phase.

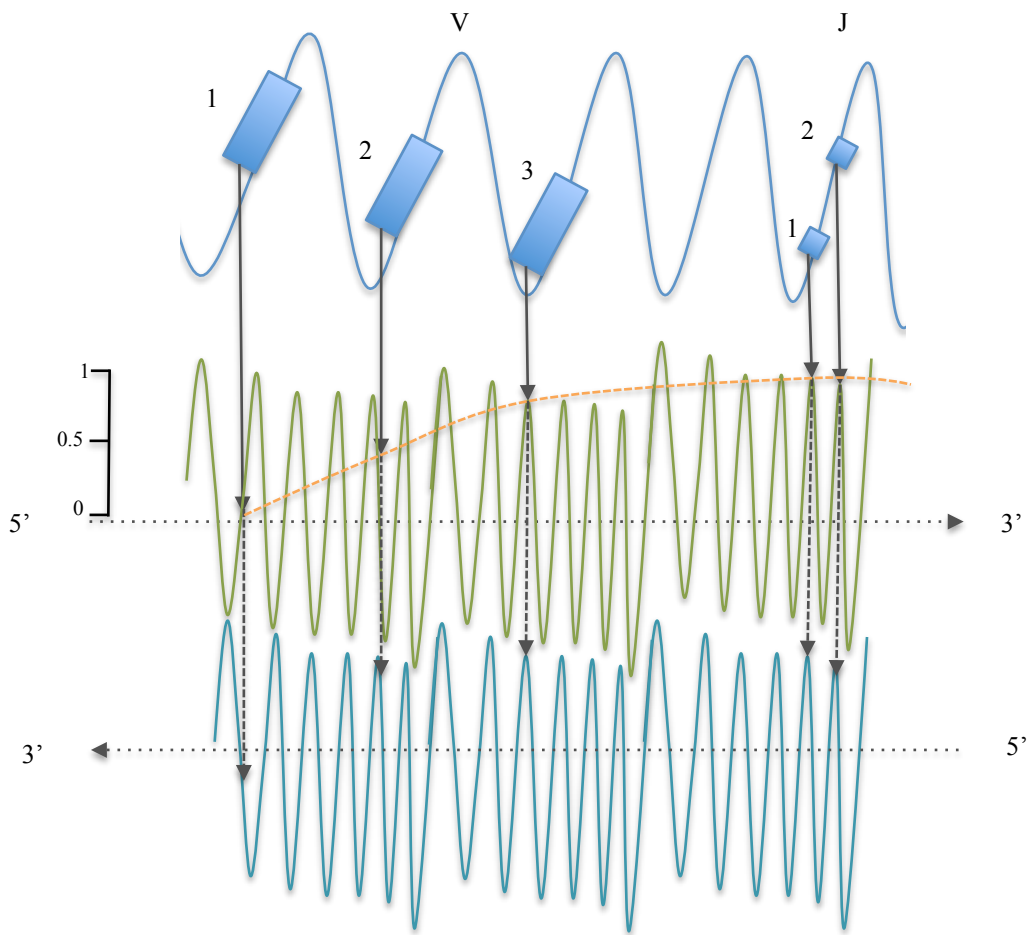

**Supplementary Table 1.** TRB locus. Source: http://www.ncbi.nlm.nih.gov/nuccore/114841177?report=graph

| Gene segment | Size of gene segment | Start position | End postion | Relative position from locus end | Spacing between segments |
|---|---|---|---|---|---|
| V1* | 450 | 91557 | 92,006 | 575,783 | N/A |
| V2 | 435 | 93,552 | 93,986 | 573,788 | 1,546 |
| V3-1 | 460 | 101,150 | 101,609 | 566190 | 7,164 |
| V4-1 | 454 | 105,774 | 106,227 | 561566 | 4,165 |
| V5-1 | 470 | 113,622 | 114,091 | 553718 | 7,395 |
| V6-1 | 433 | 120,903 | 121,335 | 546437 | 6,812 |
| V7-1* | 489 | 124,764 | 125,252 | 542576 | 3,429 |
| V4-2 | 454 | 138,078 | 138,531 | 529262 | 12,826 |
| V6-2 | 433 | 141,898 | 142,330 | 525442 | 3,367 |
| V3-2* | 460 | 151,879 | 152,338 | 515461 | 9,549 |
| V4-3 | 454 | 155,338 | 155,791 | 512002 | 3,000 |
| V6-3 | 433 | 163,579 | 164,011 | 503761 | 7,788 |
| V7-2 | 497 | 167,212 | 167,708 | 500128 | 3,201 |
| V8-1* | 287 | 173,318 | 173,604 | 494022 | 5,610 |
| V5-2* | 441 | 187,166 | 187,606 | 480174 | 13,562 |
| V6-4 | 435 | 195,522 | 195,956 | 471818 | 7,916 |
| V7-3 | 457 | 199,093 | 199,549 | 468247 | 3,137 |
| V8-2* | 461 | 200,895 | 201,355 | 466445 | 1,346 |
| V5-3* | 467 | 203,911 | 204,377 | 463429 | 2,556 |
| V9 | 475 | 206,647 | 207,121 | 460693 | 2,270 |
| V10-1 | 450 | 214,638 | 215,087 | 452702 | 7,517 |
| V11-1 | 478 | 222,397 | 222,874 | 444943 | 7,310 |
| V12-1* | 443 | 229,930 | 230,372 | 437410 | 7,056 |
| V10-2 | 450 | 239,704 | 240,153 | 427636 | 9,332 |
| V11-2 | 439 | 248,664 | 249,102 | 418676 | 8,511 |
| V12-2* | 443 | 255,591 | 256,033 | 411749 | 6,489 |
| V6-5 | 436 | 265,721 | 266,156 | 401619 | 9,688 |
| V7-4 | 462 | 269,879 | 270,340 | 397461 | 3,723 |
| V5-4 | 465 | 277,826 | 278,290 | 389514 | 7,486 |
| V6-6 | 433 | 284,296 | 284,728 | 383044 | 6,006 |
| V7-5* | 472 | 288,811 | 289,282 | 378529 | 4,083 |
| V5-5 | 465 | 297,271 | 297,735 | 370069 | 7,989 |
| V6-7* | 433 | 302,579 | 303,011 | 364761 | 4,844 |
| V7-6 | 493 | 306,894 | 307,386 | 360446 | 3,883 |
| V5-6 | 466 | 314,783 | 315,248 | 352557 | 7,397 |
| V6-8 | 429 | 322,095 | 322,523 | 345245 | 6,847 |
| V7-7 | 502 | 326,337 | 326,838 | 341003 | 3,814 |
| V5-7 | 467 | 334,800 | 335,266 | 332540 | 7,962 |
| V6-9 | 432 | 342,106 | 342,537 | 325234 | 6,840 |
| V7-8 | 484 | 346,720 | 347,203 | 320620 | 4,183 |
| V5-8 | 467 | 354,936 | 355,402 | 312404 | 7,733 |
| V7-9 | 473 | 364,137 | 364,609 | 303203 | 8,735 |
| V13 | 484 | 370,656 | 371,139 | 296684 | 6,047 |
| V10-3 | 450 | 379,068 | 379,517 | 288272 | 7,929 |
| V11-3 | 438 | 389,716 | 390,153 | 277624 | 10,199 |
| V12-3 | 447 | 395,318 | 395,764 | 272022 | 5,165 |
| V12-4 | 447 | 398,641 | 399,087 | 268699 | 2,877 |
| V12-5 | 447 | 415,832 | 416,278 | 251508 | 16,745 |
| V14 | 433 | 422,775 | 423,207 | 244565 | 6,497 |
| V15 | 469 | 427,850 | 428,318 | 239490 | 4,643 |
| V16* | 454 | 432,859 | 433,312 | 234481 | 4,541 |
| V17 | 735 | 436,475 | 437,209 | 230865 | 3,163 |
| V18 | 619 | 450,546 | 451,164 | 216794 | 13,337 |
| V19 | 476 | 453,806 | 454,281 | 213534 | 2,642 |
| V20-1 | 673 | 461,476 | 462,148 | 205864 | 7,195 |
| V21-1* | 468 | 471,659 | 472,126 | 195681 | 9,511 |
| V22-1* | 451 | 476,480 | 476,930 | 190860 | 4,354 |
| V23-1* | 504 | 480,699 | 481,202 | 186641 | 3,769 |
| V24-1 | 477 | 490,040 | 490,516 | 177300 | 8,838 |
| V25-1 | 468 | 504,415 | 504,882 | 162925 | 13,899 |
| VA* | 455 | 515,043 | 515,497 | 152297 | 10,161 |
| V26* | 485 | 529,336 | 529,820 | 138004 | 13,839 |
| V8* | 541 | 545,032 | 545,572 | 122308 | 15,212 |
| V27 | 473 | 549,038 | 549,510 | 118302 | 3,466 |
| V28 | 481 | 554,328 | 554,808 | 113012 | 4,818 |
| V29-1 | 614 | 573,946 | 574,559 | 93394 | 19,138 |
| D1 | 12 | 640,268 | 640,279 | 27,072 | 65,709 |
| J1-1 | 48 | 640,935 | 640,982 | 26,405 | 656 |
| J1-2 | 48 | 641,072 | 641,119 | 26,268 | 90 |
| J1-3 | 50 | 641,685 | 641,734 | 25,655 | 566 |
| J1-4 | 51 | 642,280 | 642,330 | 25,060 | 546 |
| J1-5 | 50 | 642,553 | 642,602 | 24,787 | 223 |
| J1-6 | 53 | 643,043 | 643,095 | 24,297 | 441 |
| C1 | 1,448 | 645,749 | 647,196 | 21,591 | 2,654 |
| D2 | 16 | 649,760 | 649,775 | 17,580 | 2,564 |
| J2-1 | 50 | 650,420 | 650,469 | 16,920 | 645 |
| J2-2 | 51 | 650,615 | 650,665 | 16,725 | 146 |
| J2-2P* | 46 | 650,752 | 650,797 | 16,588 | 87 |
| J2-3 | 49 | 650,902 | 650,950 | 16,438 | 105 |
| J2-4 | 50 | 651,053 | 651,102 | 16,287 | 103 |
| J2-5 | 120 | 651,174 | 651,293 | 16,166 | 72 |
| J2-6 | 81 | 651,266 | 651,346 | 16,074 | -27 |
| J2-7 | 47 | 651,511 | 651,557 | 15,829 | 165 |
| C2 | 1,489 | 655,095 | 656,583 | 12,245 | 3,538 |
| V30 | 701 | 666,640 | 667,340 | 700 | 10,057 |

**Supplementary table 2.** TRA locus. Source: http://www.ncbi.nlm.nih.gov/nuccore/99345462?report=graph

| Gene segment | Size of gene segment | Start position | End position | Relative position from locus end | Spacing between segments |
|---|---|---|---|---|---|
| V1-1 | 703 | 145,152 | 145,854 | 931,014 | N/A |
| V1-2 | 614 | 166,288 | 166,901 | 909,878 | 20,434 |
| V2 | 513 | 235,649 | 236,161 | 840,517 | 68,748 |
| V3 | 433 | 247,231 | 247,663 | 828,935 | 11,070 |
| V4 | 764 | 259,596 | 260,359 | 816,570 | 11,933 |
| V5 | 516 | 272,569 | 273,084 | 803,597 | 12,210 |
| V6 | 528 | 291,888 | 292,415 | 784,278 | 18,804 |
| V7 | 511 | 306,305 | 306,815 | 769,861 | 13,890 |
| V8-1 | 468 | 320,686 | 321,153 | 755,480 | 13,871 |
| V9-1 | 637 | 334,647 | 335,283 | 741,519 | 13,494 |
| V10 | 568 | 348,767 | 349,334 | 727,399 | 13,484 |
| V11* | 532 | 352,787 | 353,318 | 723,379 | 3,453 |
| V12-1 | 532 | 364,520 | 365,051 | 711,646 | 11,202 |
| V8-2 | 460 | 370,039 | 370,498 | 706,127 | 4,988 |
| V8-3 | 449 | 375,830 | 376,278 | 700,336 | 5,332 |
| V13-1 | 504 | 392,138 | 392,641 | 684,028 | 15,860 |
| V12-2 | 544 | 411,234 | 411,777 | 664,932 | 18,593 |
| V8-4 | 477 | 417,836 | 418,312 | 658,330 | 6,059 |
| V8-5* | 1,522 | 426,352 | 427,873 | 649,814 | 8,040 |
| V13-2 | 499 | 441,528 | 442,026 | 634,638 | 13,655 |
| V14DV4 | 514 | 447,409 | 447,922 | 628,757 | 5,383 |
| V9-2 | 480 | 464,464 | 464,943 | 611,702 | 16,542 |
| V15* | 544 | 473,429 | 473,972 | 602,737 | 8,486 |
| V12-3 | 555 | 488,831 | 489,385 | 587,335 | 14,859 |
| V8-6 | 442 | 502,014 | 502,455 | 574,152 | 12,629 |
| V16 | 443 | 513,830 | 514,272 | 562,336 | 11,375 |
| V17 | 483 | 521,020 | 521,502 | 555,146 | 6,748 |
| V18 | 463 | 526,545 | 527,007 | 549,621 | 5,043 |
| V19 | 553 | 530,963 | 531,515 | 545,203 | 3,956 |
| V20 | 491 | 564,004 | 564,494 | 512,162 | 32,489 |
| V21 | 536 | 575,875 | 576,410 | 500,291 | 11,381 |
| V22 | 519 | 594,048 | 594,566 | 482,118 | 17,638 |
| V23DV6 | 530 | 609,824 | 610,333 | 466,342 | 15,258 |
| DV1 | 570 | 619,422 | 619,991 | 371,242 | 9,089 |
| V24 | 504 | 628,715 | 629,218 | 447,451 | 8,724 |
| V25 | 609 | 635,471 | 636,079 | 440,695 | 6,253 |
| V26-1 | 759 | 646,580 | 647,338 | 429,586 | 10,501 |
| V8-7 | 482 | 655,610 | 656,091 | 420,556 | 8,272 |
| V27 | 542 | 671,141 | 671,682 | 405,025 | 15,050 |
| V28* | 560 | 678,107 | 678,666 | 398,059 | 6,425 |
| V30 | 560 | 691,420 | 691,979 | 384,746 | 12,754 |
| V31* | 592 | 700,265 | 700,856 | 375,901 | 8,286 |
| V32* | 500 | 708,553 | 709,052 | 367,613 | 7,697 |
| V33* | 557 | 713,147 | 713,703 | 363,019 | 4,095 |
| V26-2 | 786 | 725,572 | 726,357 | 350,594 | 11,869 |
| V34 | 596 | 730,525 | 731,120 | 345,641 | 4,168 |
| V35 | 580 | 744,887 | 745,466 | 331,279 | 13,767 |
| V37* | 1,425 | 788,727 | 790,151 | 287,439 | 43,261 |
| V38-1 | 593 | 794,949 | 795,541 | 281,217 | 4,798 |
| V39 | 500 | 827,034 | 827,533 | 249,132 | 31,493 |
| V40 | 430 | 838,017 | 838,446 | 238,149 | 10,484 |
| V41 | 504 | 843,715 | 844,218 | 232,451 | 5,269 |
| DV2 | 497 | 946,632 | 947,128 | 44,032 | 102,414 |
| DD1 | 8 | 962,634 | 962,641 | 28,030 | 15,506 |
| DD2 | 9 | 963,094 | 963,102 | 27,570 | 453 |
| DD3 | 13 | 973,200 | 973,212 | 17,464 | 10,098 |
| DJ1 | 51 | 974,176 | 974,226 | 16,488 | 964 |
| DJ4 | 48 | 979,336 | 979,383 | 11,328 | 5,110 |
| DJ2 | 54 | 980,776 | 980,829 | 9,888 | 1,393 |
| DJ3 | 59 | 983,185 | 983,243 | 7,479 | 2,356 |
| DC | 3,664 | 987,001 | 990,664 | 3,663 | 3,758 |
| DV3 | 574 | 993,128 | 993,701 | 2,464 | 2,464 |
| J61 | 60 | 999,401 | 999,460 | 76,765 | 5,700 |
| J60* | 57 | 1,000,391 | 1,000,447 | 75,775 | 931 |
| J59 | 54 | 1,000,638 | 1,000,691 | 75,528 | 191 |
| J58 | 63 | 1,001,791 | 1,001,853 | 74,375 | 1,100 |
| J57 | 63 | 1,002,956 | 1,003,018 | 73,210 | 1,103 |
| J56 | 62 | 1,003,605 | 1,003,666 | 72,561 | 587 |
| J55* | 57 | 1,005,781 | 1,005,837 | 70,385 | 2,115 |
| J54 | 60 | 1,006,371 | 1,006,430 | 69,795 | 534 |
| J53 | 66 | 1,007,088 | 1,007,153 | 69,078 | 658 |
| J52 | 69 | 1,010,311 | 1,010,379 | 65,855 | 3,158 |
| J51 | 63 | 1,011,266 | 1,011,328 | 64,900 | 887 |
| J50 | 60 | 1,012,676 | 1,012,735 | 63,490 | 1,348 |
| J49 | 56 | 1,013,571 | 1,013,626 | 62,595 | 836 |
| J48 | 63 | 1,014,574 | 1,014,636 | 61,592 | 948 |
| J47 | 57 | 1,016,933 | 1,016,989 | 59,233 | 2,297 |
| J46 | 63 | 1,017,484 | 1,017,546 | 58,682 | 495 |
| J45 | 66 | 1,018,006 | 1,018,071 | 58,160 | 460 |
| J44 | 63 | 1,018,901 | 1,018,963 | 57,265 | 830 |
| J43 | 54 | 1,019,991 | 1,020,044 | 56,175 | 1,028 |
| J42 | 66 | 1,020,967 | 1,021,032 | 55,199 | 923 |
| J41 | 62 | 1,021,737 | 1,021,798 | 54,429 | 705 |
| J40 | 61 | 1,023,767 | 1,023,827 | 52,399 | 1,969 |
| J39 | 63 | 1,025,680 | 1,025,742 | 50,486 | 1,853 |
| J38 | 62 | 1,026,310 | 1,026,371 | 49,856 | 568 |
| J37 | 62 | 1,027,830 | 1,027,891 | 48,336 | 1,459 |
| J36 | 59 | 1,029,190 | 1,029,248 | 46,976 | 1,299 |
| J35 | 59 | 1,030,722 | 1,030,780 | 45,444 | 1,474 |
| J34 | 58 | 1,031,744 | 1,031,801 | 44,422 | 964 |
| J33 | 57 | 1,032,680 | 1,032,736 | 43,486 | 879 |
| J32 | 66 | 1,033,418 | 1,033,483 | 42,748 | 682 |
| J31 | 57 | 1,035,044 | 1,035,100 | 41,122 | 1,561 |
| J30 | 57 | 1,036,927 | 1,036,983 | 39,239 | 1,827 |
| J29 | 60 | 1,038,014 | 1,038,073 | 38,152 | 1,031 |
| J28 | 66 | 1,039,694 | 1,039,759 | 36,472 | 1,621 |
| J27 | 59 | 1,040,344 | 1,040,402 | 35,822 | 585 |
| J26 | 60 | 1,042,517 | 1,042,576 | 33,649 | 2,115 |
| J25 | 60 | 1,042,883 | 1,042,942 | 33,283 | 307 |
| J24 | 63 | 1,044,040 | 1,044,102 | 32,126 | 1,098 |
| J23 | 63 | 1,044,487 | 1,044,549 | 31,679 | 385 |
| J22 | 63 | 1,046,109 | 1,046,171 | 30,057 | 1,560 |
| J21 | 55 | 1,047,666 | 1,047,720 | 28,500 | 1,495 |
| J20 | 57 | 1,048,389 | 1,048,445 | 27,777 | 669 |
| J19 | 60 | 1,049,326 | 1,049,385 | 26,840 | 881 |
| J18 | 66 | 1,049,713 | 1,049,778 | 26,453 | 328 |
| J17 | 63 | 1,050,905 | 1,050,967 | 25,261 | 1,127 |
| J16 | 60 | 1,052,580 | 1,052,639 | 23,586 | 1,613 |
| J15 | 60 | 1,053,673 | 1,053,732 | 22,493 | 1,034 |
| J14 | 52 | 1,054,369 | 1,054,420 | 21,797 | 637 |
| J13 | 63 | 1,055,117 | 1,055,179 | 21,049 | 697 |
| J12 | 60 | 1,055,980 | 1,056,039 | 20,186 | 801 |
| J11 | 60 | 1,056,543 | 1,056,602 | 19,623 | 504 |
| J10 | 64 | 1,057,536 | 1,057,599 | 18,630 | 934 |
| J9 | 61 | 1,059,593 | 1,059,653 | 16,573 | 1,994 |
| J8 | 60 | 1,060,183 | 1,060,242 | 15,983 | 530 |
| J7 | 59 | 1,061,658 | 1,061,716 | 14,508 | 1,416 |
| J6 | 62 | 1,063,167 | 1,063,168 | 13,059 | 1,391 |
| J5 | 60 | 1,064,281 | 1,064,340 | 11,885 | 1,113 |
| J4 | 63 | 1,066,233 | 1,066,295 | 9,933 | 1,893 |
| J3 | 62 | 1,067,213 | 1,067,274 | 8,953 | 918 |
| J2 | 62 | 1,068,106 | 1,068,171 | 8,060 | 832 |
| J1 | 62 | 1,069,072 | 1,069,133 | 7,094 | 901 |
| C | 4,637 | 1,071,538 | 1,076,174 | 4,628 | 2,405 |

**Supplementary Figure 1**. DNA helix as seen end on (x-axis coming out of the page), illustrating the concept of application of trigonometric ratios to nucleotide positions (*x*), representing arc-lengths on the gene locus. The angle *ac* ($2\pi x/10.4$) in radians is used to calculate the ratios, which are extrapolated to form the *sine wave* depicted below (not to scale). Preceding nucleotide position (*x-6*) is included for reference: *a* is the radius of the circle and remains uniform, lines *b* and *c* oscillate on their respective axes, imparting values ranging between +1 and -1 for the sine and cosine functions of *x* as it takes on successively larger values (angular distance from origin, in radians) as the DNA helix advances.

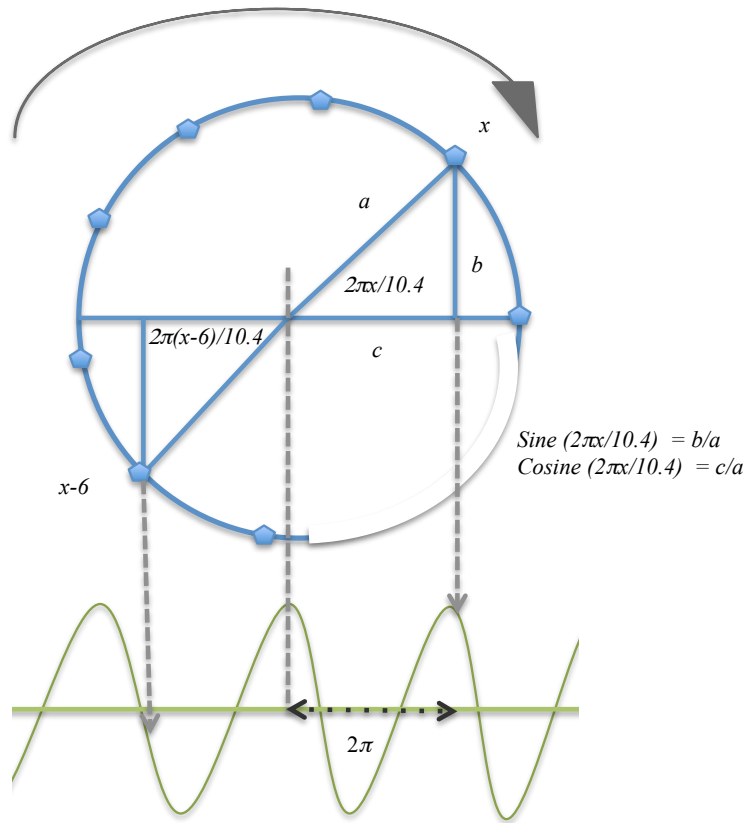

**Supplementary figure 2.** TRB V clonal frequency measured by high-throughput sequencing of cDNA isolated from circulating CD3+ cells from 6 normal stem cell donors

# References.

.